\documentclass[aps,prb,twocolumn,floatfix,superscriptaddress,showpacs,footinbib]{revtex4}
\usepackage{amsmath,amssymb,natbib,bm,graphicx,url,epsfig,color}
\usepackage[ansinew]{inputenc}

\begin{document}

\vspace*{0.5ex}

\title{Transport through anisotropic magnetic molecules with partially
ferromagnetic leads: Spin-charge conversion and negative differential
conductance}

\author{Florian Elste}
\email{felste@physik.fu-berlin.de}
\affiliation{Institut f\"ur Theoretische Physik, Freie Universit\"at Berlin,
Arnimallee 14, D-14195 Berlin, Germany}
\author{Carsten Timm}
\affiliation{Department of Physics and Astronomy, University of Kansas,
Lawrence, Kansas 66045, USA}

\date{January 13, 2006}

\begin{abstract}
We theoretically investigate inelastic transport through anisotropic
magnetic molecules weakly coupled to one ferromagnetic and one nonmagnetic
lead. We find that the current is suppressed over wide voltage ranges due to
spin blockade. In this system, spin blockade is associated with successive
spin flips of the molecular spin and depends on the anisotropy energy
barrier. This leads to the appearance of a window of bias voltages between
the Coulomb blockade and spin blockade regimes where the current is large
and to negative differential conductance. Remarkably,
negative differential conductance is also present close to room
temperature. Spin-blockade behavior is accompanied by super-Poissonian shot
noise, like in nonmagnetic quantum dots. Finally, we show that the
\emph{charge} transmitted through the molecule between initial preparation
in a certain spin state and infinite time strongly depends on the
initial spin state in certain parameter ranges. Thus the molecule can act as
a spin-charge converter, an effect potentially useful as a read-out
mechanism for molecular spintronics.
\end{abstract}


\maketitle

\section{Introduction}

The remarkable miniaturization in semiconductor-based microelectronics in the
past decades rapidly approaches its natural limits.
In this context,
electronic transport through single molecules has received much 
attention.\cite{Joachim,Emberly,Nitzan,Xue,Mitra,Reed,Park,Weber,Reichert,%
JPark,Liang,Koch,Heersche,Elste,Romeike1,Timm,Romeike2} Since individual
molecules are about two orders of magnitude smaller than chip features
presently attainable, devices composed of single molecules would present an
important breakthrough. Besides, transport through single molecules allows one to study
nonequilibrium processes in systems dominated by quantum effects, but having
a limited number of relevant degrees of freedom. 
This is particularly interesting in the presence of vibrations and local magnetic moments.
Transport through single molecules is thus of interest for fundamental physics as well.

It is very promising to combine molecular electronics with
spintronics, i.e., the idea to employ spins to store and process
information, in order to build more efficient memory devices on the 
nanometer scale.\cite{Zutic,Wolf}
The combination of these two strategies could be realized with 
magnetic molecules, i.e., molecules with a local 
magnetic moment.\cite{Blundell}

Experimental work in this field has focussed on the Zeeman splitting of 
the Coulomb-blockade (CB) peaks as well as the Kondo effect.\cite{JPark,Liang} 
More recent transport measurements for a $\mathrm{Mn_{12}}$ derivative have exhibited 
fine structure of the CB peaks originating from \textit{magnetic} excitations and regions of 
complete current suppression due to spin selection rules.\cite{Heersche}
These experiments have also led to a number of
theoretical works.\cite{Heersche,Elste,Romeike1,Timm,Romeike2}
Furthermore, transport through quantum dots coupled to magnetic leads has 
been studied extensively.\cite{Koenig,Braun,Cottet2,Braig}
In particular, it has been shown for pa\-ra\-mag\-ne\-tic\cite{Cottet1}
and ferromagnetic\cite{Cottet2} leads that
an external magnetic field, which lifts the spin degeneracy of energy levels, leads to
voltage ranges above the CB threshold where the current is strongly suppressed due to
different tunneling rates for spin-up and spin-down electrons.\cite{Cottet1}
However, magnetic molecules coupled to magnetic leads have
received little attention.

An essential requirement for spintronics is the ability to effectively control
and detect the spin. In a recent paper we have shown  that magnetic
\textit{anisotropy} is crucial for slow spin relaxation in magnetic molecules
and can lead to giant spin amplification:\cite{Timm} If the molecule is
prepared in a magnetic initial state, the current can be highly polarized for
an exponentially long time interval. Thus the spin moment transmitted through
the molecule depends on the initial orientation of the molecular spin and can
be much larger than the initial molecular spin itself.\cite{Timm} This allows
one to effectively \textit{read out} the spin of a molecule coupled to two
nonmagnetic leads.
The strong anisotropy is expected to arise from the interplay of the ligand
field (e.g., in porphyrin complexes or $\mathrm{Mn_{12}}$
derivatives\cite{Heersche}) and spin-orbit coupling as well as from the
interaction with image dipoles.

In the present paper we study the inelastic transport through an anisotropic
magnetic molecule weakly coupled to one nonmagnetic and one
\textit{ferromagnetic} lead employing the rate-equation
approach.\cite{Mitra,Koch,Heersche,Elste,Timm}
This configuration is motivated by the possibility to switch the
molecule to a predetermined spin state, i.e., to \textit{write} the spin, by
applying a bias voltage alone, in zero magnetic field.\cite{Timm} 
We show that the proposed configuration leads to interesting physics beyond the effect
of spin writing, including the occurrence of large negative differential 
conductance (NDC) at \textit{high} temperatures. This effect is distinct from
the NDC found  at
\emph{low} temperatures in the fine structure of differential-conductance
peaks due to inelastic processes.\cite{Heersche,Timm} In the low-temperature
transport the anisotropy leads to the appearance of a finite window of bias
voltages for which the current is large, whereas it is strongly suppressed
on either side due to CB and \emph{spin blockade}
(SB),\cite{Heersche,Weinmann,Imamura,Romeike1,Huettel,Ciorga,Cottet1,Cottet2}
respectively. By SB we mean the suppression of the current due
to small single-electron tunneling rates, in our case due to
density-of-states effects. In our system the large local spin leads to
interesting modifications of SB, as discussed below.

We also find that the \emph{charge} transmitted through a molecule prepared
in a particular spin state depends strongly on this initial state in certain
parameter regimes. In fact the difference of the total transmitted charge
between preparation and time $t\to\infty$ is an \emph{exponential} function
of a certain energy difference over temperature. This effect is related to
the giant spin amplification mentioned above,\cite{Timm} but appears
in the charge channel.

\section{Model and Methods}

Our results are obtained for a molecule weakly coupled to two metallic
leads. Relaxation in the leads is assumed to be sufficiently fast so that
their electron distributions can be described by equilibrium Fermi
functions. We assume transport to be dominated by sequential tunneling
through a single molecular level with onsite energy $\varepsilon$ and local
Coulomb repulsion $U$. The full Hamiltonian of the system reads $H =
H_{\text{mol}} + H_{\text{leads}} + H_{\text{t}}$, where\cite{Timm}
\begin{eqnarray}
H_{\text{mol}} & = & \left(\epsilon -e V_{\text{g}}\right)n
  + \frac{U}{2}n\left(n-1\right)-J\,\mathbf{s}\cdot\mathbf{S} -K_2(S^z)^2
  \nonumber \\
& & {} -B\left(s^z+S^z\right) \label{Hamiltonian}
\end{eqnarray}
describes the molecular degrees of freedom,
$H_{\text{leads}} = \sum_{\alpha=\text{L},\text{R}} \sum_{\mathbf{k} \sigma}
  \epsilon_{\alpha \mathbf{k}}
  a_{\alpha \mathbf{k} \sigma}^{\dagger} a_{\alpha \mathbf{k} \sigma}$
represents the two leads $\alpha=\text{L},\text{R}$ (left, right), and
$H_{\text{t}}= \sum_{\alpha=\text{L},\text{R}} \sum_{n \mathbf{k} \sigma}
  (t_{\alpha} a_{\alpha \mathbf{k} \sigma}^{\dagger} c_{\sigma}
  + t_{\alpha}^{\ast} c_{\sigma}^{\dagger} a_{\alpha \mathbf{k} \sigma})$
describes the tunneling.
Here, the operator $c_{\sigma}^{\dagger}$ creates an electron with spin
$\sigma$ on the molecule. $n_{\text{d}}
\equiv c^{\dagger}_{\uparrow}c_{\uparrow} +
c^{\dagger}_{\downarrow}c_{\downarrow}$ and $\mathbf{s} \equiv \sum_{\sigma
\sigma'} c^{\dagger}_{\sigma} \mbox{\boldmath$\sigma$}_{\sigma\sigma'}
c_{\sigma'} /2$  are the corresponding number and spin operator,
respectively. $J$ denotes
the exchange interaction between the electrons and the local spin
$\mathbf{S}$. We assume the anisotropy of $\mathbf{S}$ to be of
\emph{easy-axis} type, $K_2>0$. For simplicity we consider identical g
factors for $\mathbf{s}$ and $\mathbf{S}$. The external magnetic field is
applied along the anisotropy axis and a factor $g\mu_B$ has been absorbed
into $B$. 
Unless stated otherwise, the value of the gate voltage $V_\text{g}$ is chosen
as zero.
$a_{\alpha \mathbf{k} \sigma}^{\dagger}$
creates an electron in lead $\alpha$ with spin $\sigma$, momentum
$\mathbf{k}$ and  energy $\epsilon_{\alpha \mathbf{k}}$.   

The eigenstates of the unperturbed Hamiltonian $H_{\mathrm{mol}}$ 
fall into sectors with $n=0,1,2$ electrons\cite{Timm}. Since
$[S_{\mathrm{tot}}^z,H_{\mathrm{mol}}]=0$, the eigenvalue $m$ of
$S_{\mathrm{tot}}^z$ is a good quantum number. For $n=0$, $2$ we
obtain 
\begin{equation}
\epsilon(0,m) = -K_2 m^2 - B m
\end{equation}
and 
\begin{equation}
\epsilon(2,m) = 2(\epsilon-eV_g) + U-K_2 m^2 - B m.
\end{equation}
For $n=1$ and $-S+1/2\le m\le S-1/2$ there are 
two orthogonal states with energies
\begin{equation}
\epsilon^\pm(1,m) = \epsilon - eV_g -Bm+\frac{J}{4}
  -K_2 \Big(m^2+\frac{1}{4}\Big) \pm \Delta E(m)
\end{equation}
where $\Delta E(m) \equiv [{K_2(K_2-J)m^2+({J}/{4})^2 (2S+1)^2}]^{1/2}$.
For the fully polarized states with $n=1$ and $m=\pm(S+1/2)$  
the upper (lower) sign applies if $K_2-J/2$ is positive (negative).

Second-order perturbation theory in the hopping matrix element $t_\alpha$
produces a
\emph{Fermi's} \emph{Golden} \emph{rule} expression for the transition
rates between two many-particle states $|n\rangle$ and $|m\rangle$ of the
molecule,\cite{Mitra,Koch,Elste,Timm}
\begin{eqnarray}
R_{n\rightarrow m} & = & \sum_{\alpha=\text{L},\text{R}} \sum_{\sigma}
  \frac{2 \pi |t|^2 D^{\alpha}_{\sigma} v_{\text{uc}}}{\hbar} \Big(
  f(\epsilon_m-\epsilon_n-\mu_{\alpha})|C_{nm}^{\sigma}|^2 \nonumber \\
& & {}+ \left[1-f(\epsilon_n-\epsilon_m-\mu_{\alpha})\right]
  |C_{mn}^{\sigma}|^2 \Big) .
\label{CT.rates}
\end{eqnarray}
Here, $D^{\alpha}_{\sigma}$ denotes the density of states of electrons with
spin $\sigma$ in lead $\alpha$, $v_{\text{uc}}$ is the volume of the unit
cell, $f$ denotes the Fermi function, and the
$\epsilon_n$ are the eigenenergies of $H_{\text{mol}}$ of many-particle
eigenstates $|n\rangle$. 

The matrix elements $C_{nm}^{\sigma}$ are defined by $C_{nm}^{\sigma} \equiv \langle n|
c_{\sigma}|m \rangle$. The occupation probabilities $P^n(t)$ of
states $|n\rangle$ are obtained by solving the rate equations,
\begin{equation}
\frac{dP^n}{dt}=\sum_{m}\left(P^m R_{m\rightarrow n}-P^n
  R_{n\rightarrow m}\right).\label{rate-equations}
\end{equation}
For the steady-state probabilities the left-hand side of this equation
vanishes. In the following we assume
$\varepsilon=1\,\mathrm{eV}$ and $U=2\,\mathrm{eV}$. Then the doubly
occupied molecular states have exponentially small probabilities
at the voltages and temperatures we consider.
Also, the densities of states satisfy
$D^{\text{L}}_{\downarrow}/D^{\text{L}}_{\uparrow}=1$  and
$D^{\text{R}}_{\downarrow}/D^{\text{R}}_{\uparrow}=0.01$, i.e., lead
$\text{L}$ is nonmagnetic and lead $\text{R}$ is a nearly half-metallic
ferromagnet. For simplicity we assume
$D^{\text{L}}_\uparrow=D^{\text{R}}_\uparrow$. The local spin is chosen as
$S=2$.

The current through lead $\alpha$ reads\cite{Timm}
\begin{equation}
I^{\alpha} \equiv \mp e \sum_{m n} \left(n_n - n_m \right)
  P^m R_{m\rightarrow n}^{\alpha} \label{current},
\end{equation}
where the upper (lower) sign pertains to $\alpha=\mathrm{L}$ ($\mathrm{R}$),
$n_n$ denotes the occupation number of the eigenstate $|n\rangle$
and $R_{m\rightarrow n}^{\alpha}$ contains only those terms
in the rates, Eq.~(\ref{CT.rates}), that involve lead $\alpha$.
A rate equation approach is also used to compute the current noise spectrum of the
system,
\begin{equation}
S_{\alpha\beta}(\omega) =
  2 \int_{-\infty}^{\infty} dt\,e^{i \omega t} \langle \delta I^{\alpha}(t)\,
    \delta I^{\beta}(0) \rangle,
\label{CT.Somega}
\end{equation}
where $\delta I^{\alpha}(t)\equiv I^{\alpha}(t) - \langle I^{\alpha}\rangle$
denotes the current fluctuations in lead $\alpha$ and $\langle I^\alpha\rangle$
the steady-state current.\cite{Korotkov}

\section{Results}
\label{sec.results}

\begin{figure}[t]
\begin{center}
\hspace{1.1cm}\textbf{(a)}\hspace{3.0cm}\textbf{(b)}\\[1ex]
\includegraphics[width=8.0cm]{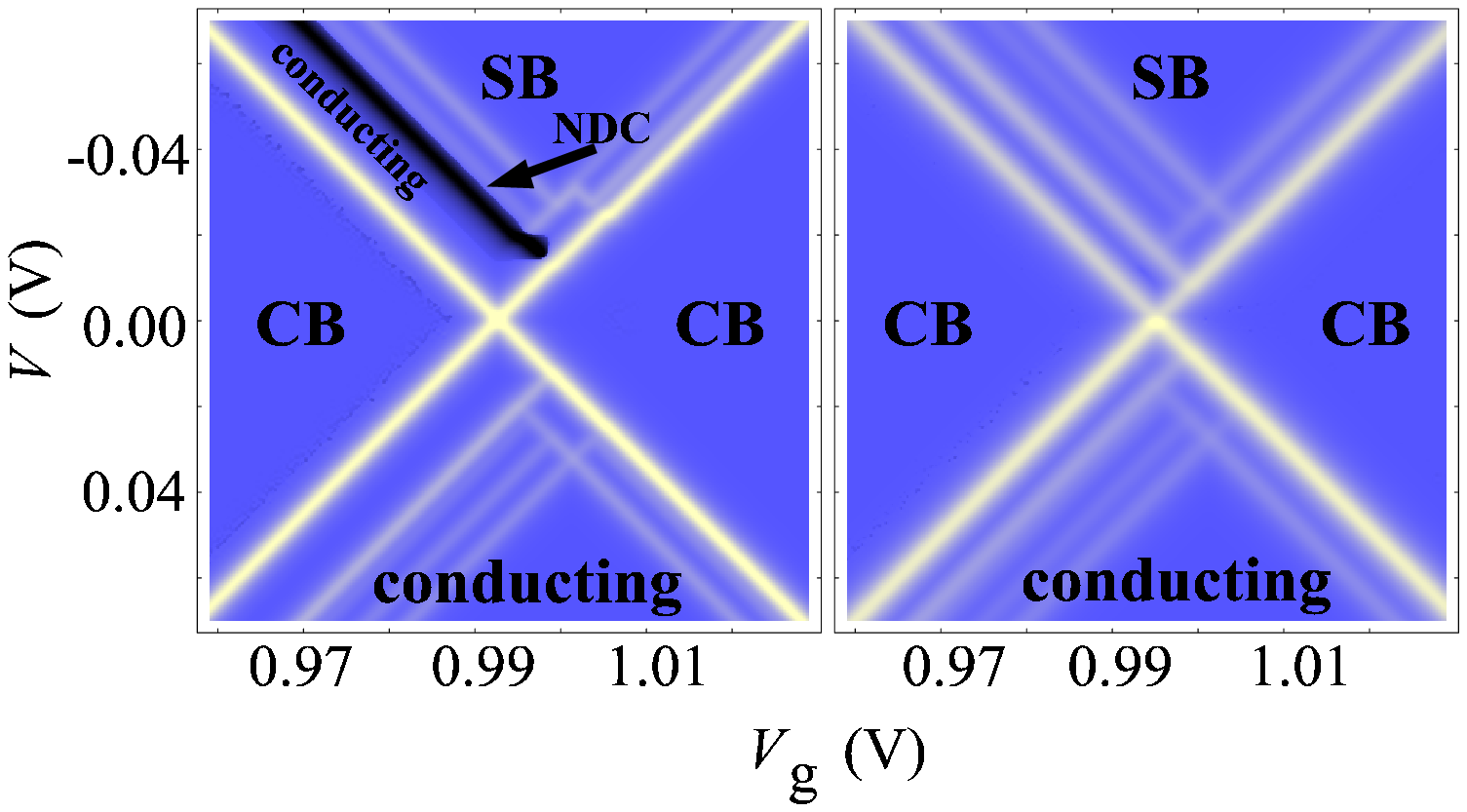}\\
\hspace{1.0cm}\textbf{(c)}\\
\hspace{1.0cm}\includegraphics[width=7.5cm]{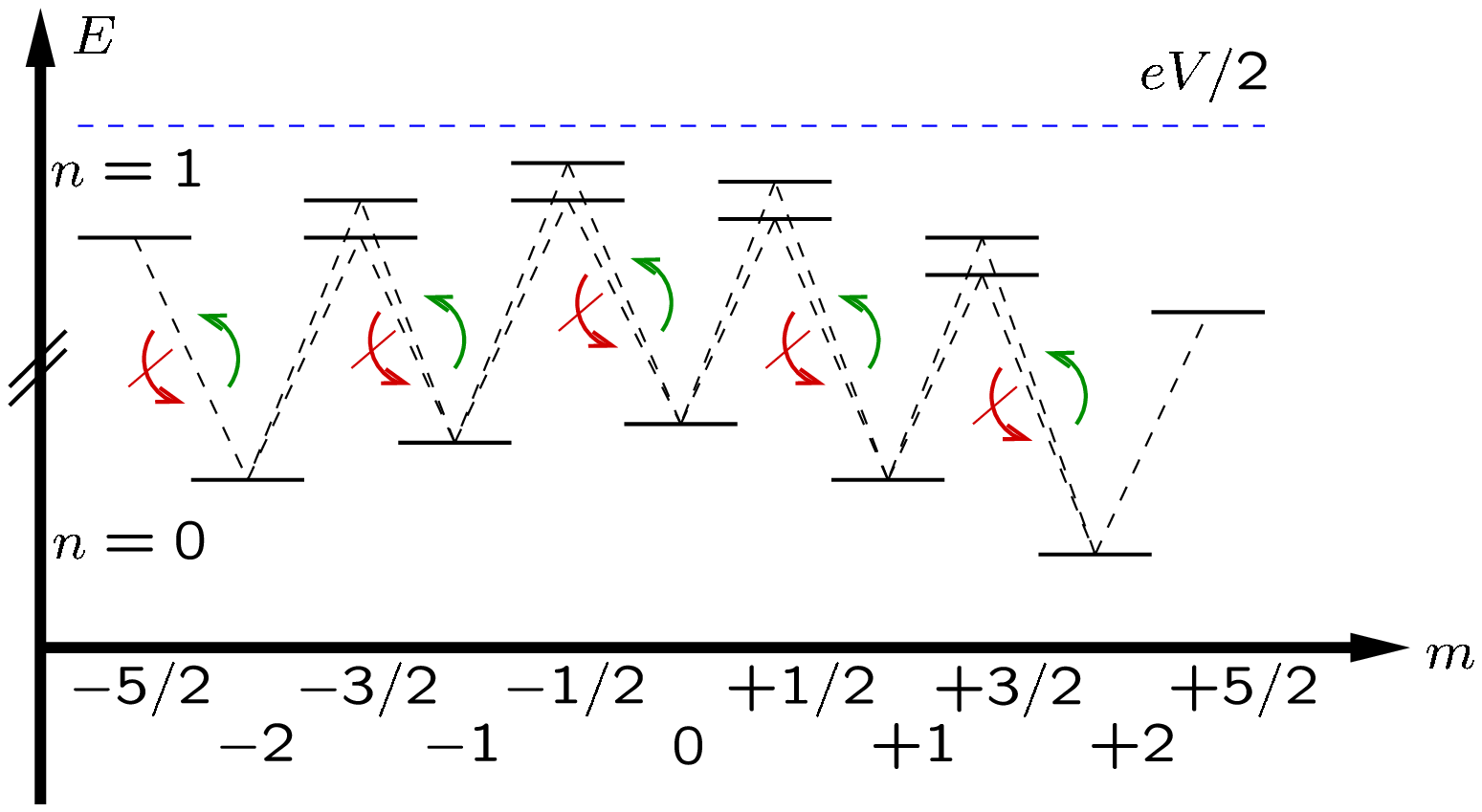} \vspace{0.5cm}
\caption{(Color online) Two-dimensional plots of the differential
conductance $dI/dV$ in the vicinity of one particular degeneracy point at
low temperatures. (a) $dI/dV$ for Zeeman splitting $B=0.05\,\mathrm{eV}$.
(b) $dI/dV$ for vanishing magnetic field. The model parameters are
$J=K_2=5\,\mathrm{meV}$.
When the magnetic field is switched on, the
CB regime and the SB regime are separated by a
finite window of bias voltage for which the conductance is high. Since it is
low (governed by $D^{\text{R}}_{\downarrow}/D^{\text{R}}_{\uparrow}$) in the
SB regime, NDC must occur where the conductance drops again. (c)
Scheme of the molecular energy levels with all allowed transitions
involving the states with $n=0$ and $n=1$ electrons in an external magnetic
field. In the SB regime all transitions are \emph{energetically} possible if
the bias voltage $V$ is large enough. However, the rates for a spin-down
electrons tunneling from the molecule into the right lead (curved arrows
with slash) are strongly suppressed due to the small density of
states.}\label{FIG1}
\end{center}
\end{figure}

The anisotropy of the local spin partially lifts the degeneracy of the
molecular energy levels with respect to the magnetic quantum number $m$,
where $m$ denotes the eigenvalues of the $z$ component of the total spin
$\textbf{S}_{\text{tot}}$.\cite{Heersche,Timm,Romeike2} This leads to a
splitting of peaks of the differential conductance $dI/dV$ at low
temperatures, as shown in Fig.~\ref{FIG1}(a) for finite and in
Fig.~\ref{FIG1}(b) for vanishing magnetic induction
$B$.\cite{rem.superscript} The complicated fine structure in the vicinity of
the degeneracy point $V_0$ arises from the anisotropy of the local spin and
the exchange  interaction of the electrons in the molecular orbital and the
local spin. Each fine structure peak corresponds to another transition
becoming available for single-electron tunneling.\cite{Heersche,Elste,Timm}
Note that $dI/dV$ is \emph{asymmetric} with respect to the bias voltage,
i.e., several fine-structure peaks have a significantly different intensity
when the bias changes sign. This results from the breaking of spin-rotation
symmetry by the ferromagnetic lead together with the spin selection rules
for the tunneling processes.

The $V$-$V_{\text{g}}$ map shows three transport regimes: At low bias the
current is thermally suppressed due to CB, except close to the degeneracy
point. The electron number on the molecule is constant, i.e., $n=0$ for
$V_{\text{g}}<V_0$ and  $n=1$ for $V_{\text{g}}>V_0$, since all transitions
between different molecular charge states are energetically forbidden. At
large \emph{positive} bias the conductivity of the molecule is high, since
the electrons in the right lead have enough energy to overcome the energy
barrier between the $n=0$ and $n=1$ states. This is the conducting regime.
At large \emph{negative} bias the conductivity of the molecule is low. This
current suppression is due to the SB mechanism explained in the
following.

Two main definitions of SB are used in the literature on transport
through quantum dots. The original definition refers to the phenomenon that
transition probabilities for single-electron tunneling vanish between states
corresponding to successive electron numbers if the total spins differ by
more than $\Delta S=1/2$.\cite{Heersche,Weinmann,Imamura,Romeike1,Huettel}
The other, more general definition refers to the situation that the
tunneling rate for electrons of one spin direction is strongly suppressed
relative to the other, e.g., due to ferromagnetic leads\cite{Ciorga,Cottet2}
or Zeeman splitting.\cite{Cottet1} In this case the system can be stuck in a
particular molecular many-body state because the rates for leaving this
state are small.

In our case the SB is related to the second
mechanism,\cite{Ciorga,Cottet1,Cottet2} but the interaction between the
electrons on the molecule and the local anisotropic spin leads to
modifications: As soon as the bias is sufficiently high, all transitions
between the $n=0$ and the $n=1$ multiplets are energetically possible, as
shown in Fig.~\ref{FIG1}(c). Selection rules for sequential tunneling
require $\Delta m = 1/2$. Both spin-up and spin-down electrons hop
\emph{onto} the molecule with equal rates, since the densities of states
$D^{\text{L}}_{\uparrow}$ and $D^{\text{L}}_{\downarrow}$ in the left
(incoming) lead are equal for both spin directions. On the other hand,
spin-up electrons leave the molecule much faster due to the polarization of
the right (outgoing) lead.  Electrons keep flowing through the molecule
until a spin-down electron tunnels in. This spin-down electron can leave the
molecule only with a very small tunneling rate due to the low density of
states $D^{\text{R}}_{\downarrow}$. On the other hand, it can rapidly leave
the molecule as a spin-\emph{up} electron if the local spin is
simultaneously reduced by unity.
However, the number of possible spin flips is limited, depending on the
initial spin state. Therefore, the molecule finally ends up in the singly
charged state with minimal spin, i.e., $S_{\text{tot}}=-S-1/2$. Further
electron tunneling is \textit{blocked}, since the left lead is energetically
unreachable and the right lead has a low density of states for spin-down
electrons, cf.\ Fig.~\ref{FIG1}(c).

\begin{figure}[t]
\begin{center}
$\begin{array}{c}
\hspace{1cm}\textbf{(a)}\\[1ex]
\includegraphics[width=7.5cm,angle=0]{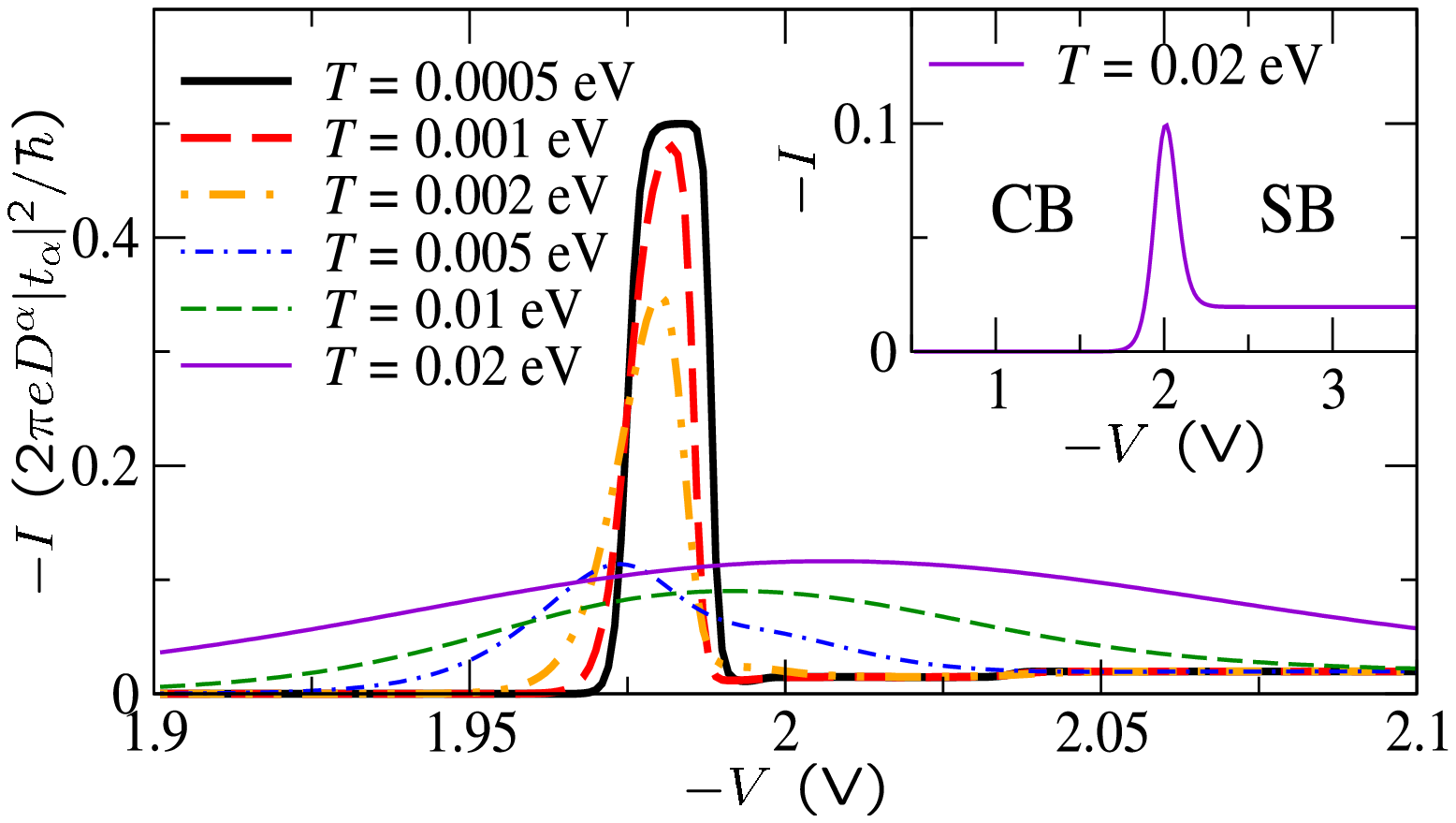}\\[1ex]
\hspace{1cm}\textbf{(b)}\\
\hspace{1cm}\includegraphics[width=7.5cm,angle=0]{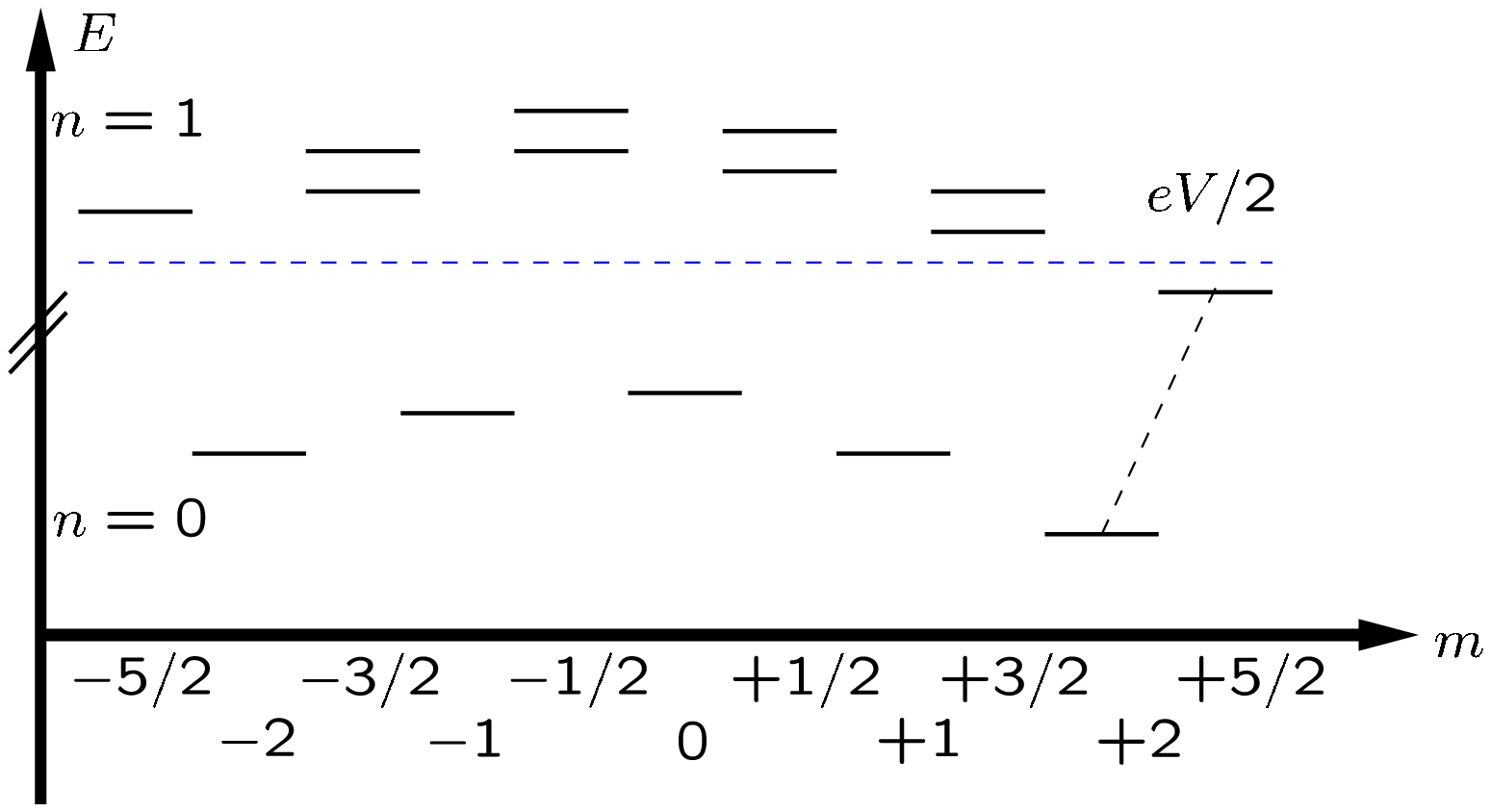}\\
\end{array}$
\vspace{0.5cm}
\caption{(Color online) (a) Current-voltage characteristics in the 
vicinity of the CB threshold assuming $K_2=J=B=0.05\,\mathrm{eV}$.
The inset shows the current for $T=0.02\,\mathrm{eV}$ over a broader
voltage range.
(b) Energy level scheme for the $n=0$ and $n=1$ spin multiplets 
as a function of the magnetic quantum number $m$.
The bias voltage is just high enough to allow 
the transition between the many-particle states with spin $m=S$ and $m=S+1/2$.
The steady-state current is highly spin-polarized, 
since the tunneling of spin-down electrons is thermally suppressed.
This regime corresponds to the \textit{plateau} with enhanced current
in (a).}\label{FIG2}
\end{center}
\end{figure}

Another interesting feature shown in Fig.~\ref{FIG1}(a) is the appearance of
a finite window of bias voltages between the CB and  SB regimes for which
the conductivity of the molecule is high. This obviously leads to NDC when
the SB regime with low conductivity is entered. Figure \ref{FIG2}(a) shows
the current-voltage characteristics  in the vicinity of the CB threshold at
zero gate voltage for finite magnetic induction. The external field tilts
the molecular energy levels
with respect to the magnetic quantum number $m$ due to
the additional Zeeman energy, as sketched in Fig.~\ref{FIG2}(b). This
removes the degeneracy of the spin multiplets. For small bias voltages and
low temperatures the current is suppressed due to CB. But as soon as the
transition from the ground state of the $n=0$ multiplet with $m=S$ to the
lowest-energy state of the $n=1$ multiplet with $m=S+1/2$ becomes
energetically allowed, cf.\ Fig.~\ref{FIG2}(b), the current increases to
the plateau shown in Fig.~\ref{FIG2}(a). These two levels are then equally
occupied. Moreover, the current through the molecule is highly
spin-polarized, since spin-up electrons tunnel rapidly through the molecule
whereas tunneling of spin-down electrons is thermally suppressed. When the
bias is further increased to allow tunneling also of spin-down electrons,
the molecule in several steps goes over to the state with $n=1$ and
$m=-S-1/2$ and the current is strongly suppressed by the SB mechanism
discussed above. The first required transition from $n=0$ and $m=S$ to $n=1$
and $m=S-1/2$ has the highest energy so that all following transitions
become active at the same bias.\cite{Elste}
The width $\Delta V$ of the window of bias voltages with
large current corresponds to twice the difference of these two
excitation energies and is given by\cite{Timm}
\begin{equation}
e\Delta V = 2\big[ 2S K_2 + B - \Delta E(S+1/2) - \Delta E(S-1/2) \big],
\end{equation}
where $\Delta E(m) \equiv [K_2(K_2-J)m^2+(J/4)^2(2S+1)^2]^{1/2}$. This holds
as long as $D^{\text{R}}_{\downarrow}/D^{\text{R}}_{\uparrow}\ll 1$ and the
Zeeman energy $B$ is not too small---for $B\to 0$ we have to take into
account that the states with $n=0$ and $m=\pm S$ become degenerate ground
states. The observation of the current plateau requires the
temperature to be small compared to $e\Delta V$, see Fig.~\ref{FIG2}(a).

\begin{figure}[t]
\begin{center}
$\begin{array}{c}
\hspace{1.0cm} \textbf{(a)}\\[3ex]
\includegraphics[width=8.0cm,angle=0]{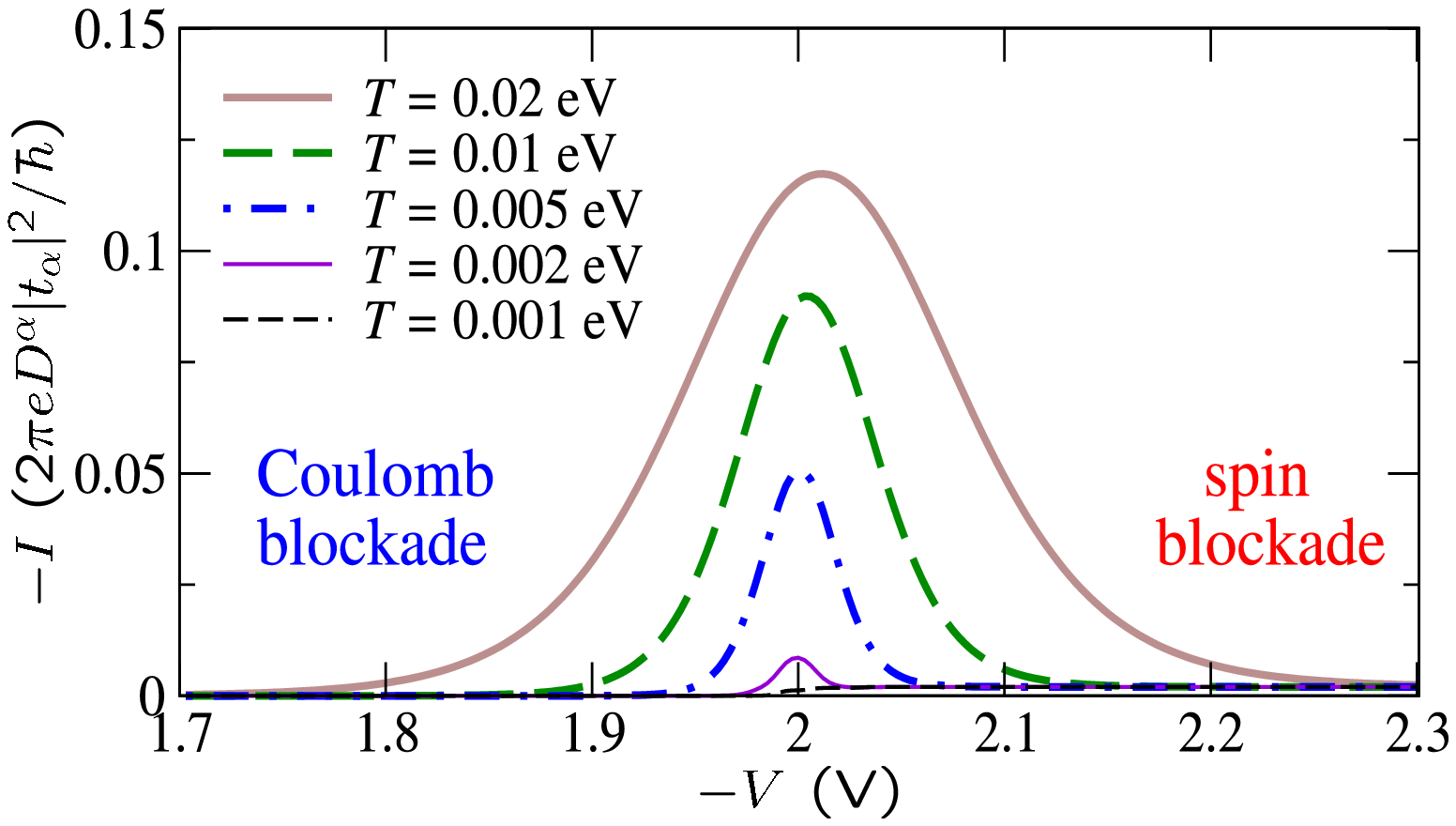}\\[1ex]
\hspace{1.0cm} \textbf{(b)}\\[2ex]
\includegraphics[width=8.0cm,angle=0]{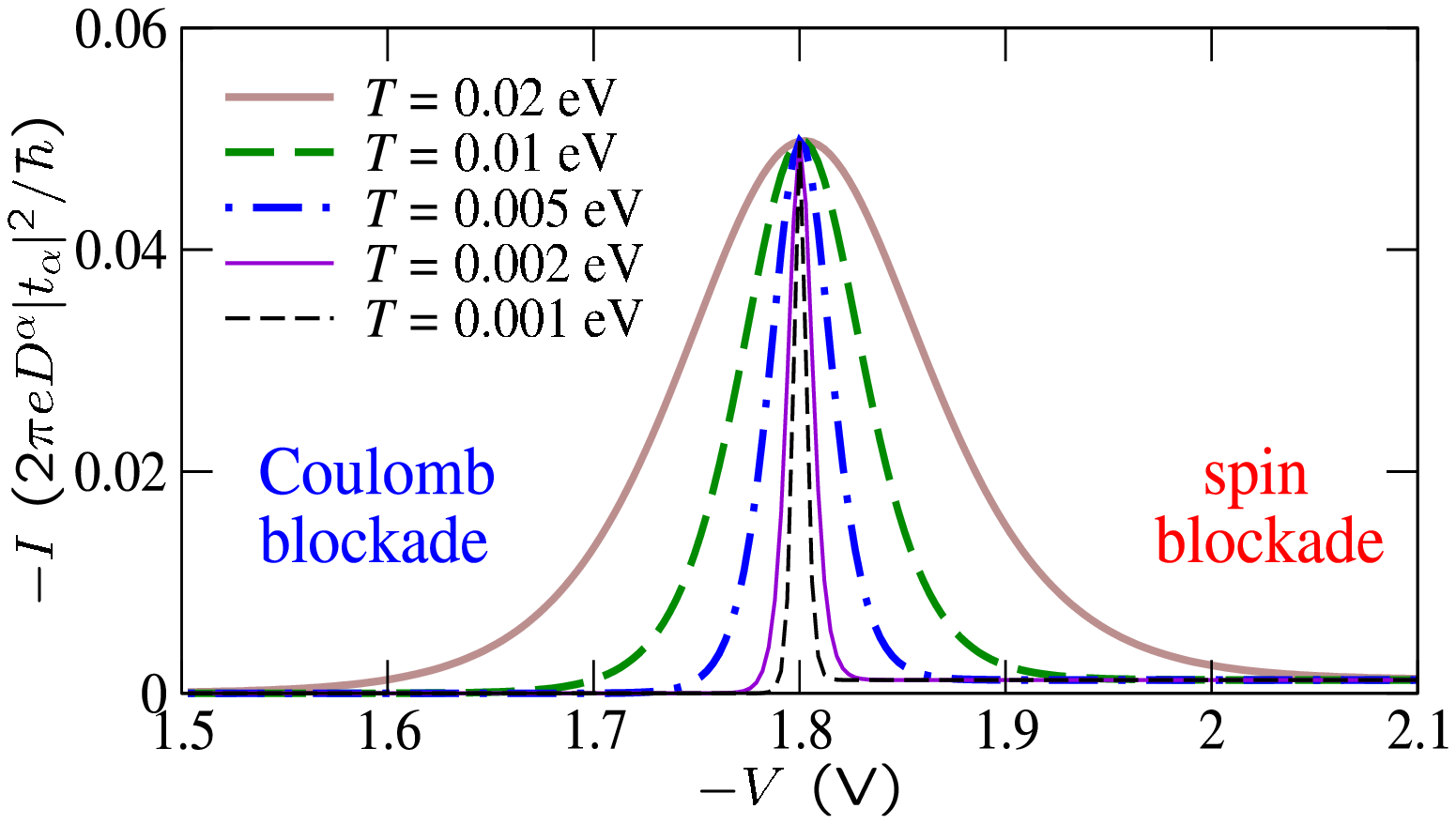}\\[1ex]
\hspace{1.0cm} \textbf{(c)}\\
\hspace{1.0cm} \includegraphics[width=7.5cm,angle=0]{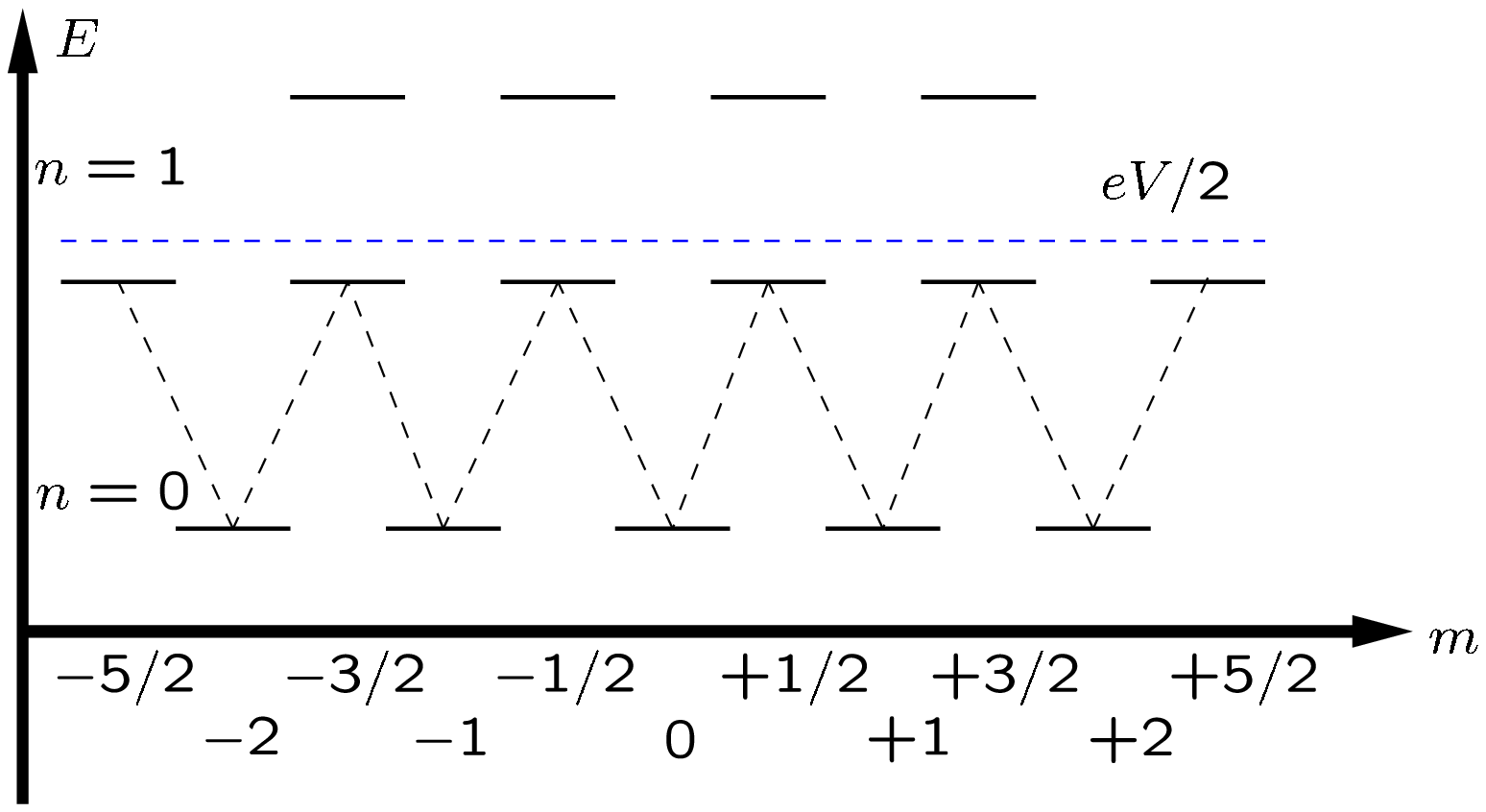}\\[2ex]
\end{array}$
\caption{(Color online) Temperature dependence of the current-voltage
characteristics at high temperatures for vanishing magnetic field,
exhibiting NDC. In (a) we assume
$K_2=J=5\,\mathrm{meV}$, corresponding to the spin multiplets schematically
shown in Figs.~\protect\ref{FIG1} and \protect\ref{FIG2}, except that now
$B=0$. In (b) we assume a large
exchange interaction $J=0.1\,\mathrm{eV}$ and vanishing magnetic
anisotropy $K_2=0$. (c) Energy level scheme for the $n=0$ and $n=1$
multiplets as a function of the magnetic quantum number $m$ for the case in
(b).}\label{FIG3}
\end{center}
\end{figure}

At \textit{high} temperatures the current steps broaden so that the plateau
of high current vanishes. Nevertheless, the current remains large at bias
voltages close to the transition point from CB to SB, as shown in the inset
in Fig.~\ref{FIG2}(a). For vanishing magnetic induction, $B=0$, there is no
enhanced current at low temperatures, as we will explain below.
Nevertheless, at high temperatures a pronounced current maximum develops
close to the transition from CB to SB, as shown in Fig.~\ref{FIG3}(a). For
anisotropy $K_2>0$, the peak grows and broadens with increasing temperature.
Since deep in the SB regime the current remains small, a broad region of NDC
develops. This is remarkable, since NDC, which has
already been observed or predicted for many systems, is usually a
low-temperature effect. The high-temperature NDC is a distinct effect, since
it appears even when there is no NDC at low temperatures.

Before we explain the high-temperature NDC, we note that the temperature
dependence of the $I$-$V$ curves is qualitatively different for vanishing
anisotropy barrier and \textit{large} exchange interaction, $K_2\ll kT \ll
J$, as shown in Fig.~\ref{FIG3}(b). Most striking is the fact that the
maximum of the peaks stays constant  for $T\rightarrow 0$. For this
situation the energy levels are shown in Fig.~\ref{FIG3}(c). It is obvious
that only a \emph{single} energy difference is relevant as long as the
higher $n=1$ quartet is not occupied. If $eV/2$ exactly equals this
transition energy, the Fermi functions in the rates, Eq.~(\ref{CT.rates}),
for these transitions all equal $1/2$ and are thus independent of
temperature. Consequently, the steady-state probabilities and current are
also independent of temperature at this bias. For lower bias all
transitions from $n=0$ to $n=1$ are thermally suppressed and for higher bias
the relevant transitions become fully active (the Fermi functions approach
unity) and we enter the SB regime discussed above. In both regimes the
current decreases for $T\to 0$.\cite{rem.maximum} In the SB regime this
leads to NDC. It is remarkable that the SB---a quantum
effect---leads to large NDC at room temperature.

For the case of $K_2>0$ and, more importantly, small $J$,
Fig.~\ref{FIG3}(a), the current at the maximum increases with temperature,
since transport through the higher $n=1$ quartet contributes more and more.
For the same reason the current maximum shifts to larger bias voltages. On
the other hand, the current maximum vanishes for $T\to 0$. In
Fig.~\ref{FIG3}(a) we have assumed vanishing magnetic induction and there is
no window of enhanced current at low temperatures: At small bias, in the CB
regime, the steady state has predominantly $n=0$ and $m=-S$ due to the
asymmetric tunneling rates into the right lead. As the bias is increased
the system \emph{directly} crosses over to the SB regime when the bias
equals the excitation energy to the state with $n=1$ and $m=-S-1/2$. 

The shape of the NDC peaks in Fig.~\ref{FIG3}(b) for vanishing anisotropy
can be fitted by a function of the form $1/\cosh^2 x$:
Coming from the CB regime,
the increase of the bias voltage energetically enables tunnelling processes
with a rate proportional to the Fermi function $f[(E-eV/2)/kT]$, 
where $E$ denotes the excitation energy. However,
the probability for the molecule being trapped in the SB state is also
proportional to $f[(E-eV/2)/kT]$. Therefore, the current also contains a factor
of the form $(1-f[(E-eV/2)/kT])+C$, where the first term accounts for the
probability of the molecule \emph{not} being trapped and the constant $C$
describes the small tunneling probability for an electron with the ``wrong''
spin. $C$ is determined by the ratio of densities of states
$D^{\mathrm{R}}_\downarrow/D^{\mathrm{R}}_\uparrow$ in the ferromagnetic lead.
Altogether, we obtain $I \sim 1/\cosh^2[(E-eV/2)/2kT] + f[(E-eV/2)/kT]C$,
which is dominated by the $1/\cosh^2 x$ term for small
$D^{\mathrm{R}}_\downarrow/D^{\mathrm{R}}_\uparrow$.
This should be compared to Ref.~\onlinecite{Bee91},
where the shape of the peaks in
the \emph{differential conductance} due the opening of additional transport
channels is considered. In our case the peak occurs in the \emph{current}.

Note that this simple argument does not apply to the 
NDC peaks shown in Fig.~\ref{FIG3}(a). For nonzero anisotropy $K_2$, the 
ground state is no longer degenerate
and the various transitions
between states of the $n=0$ and $n=1$ multiplets have different energies, which
leads to a more complex line shape.

\begin{figure}[t]
\centerline{\includegraphics[width=6.5cm]{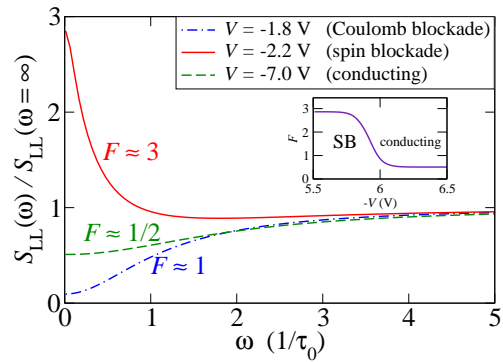}}
\caption{(Color online) Current noise spectrum. Shown are the normalized
correlation functions $S_{\text{LL}}(\omega)/S_{\text{LL}}(\omega=\infty)$
for three different bias voltages representing the three transport regimes:
$V=1.8\,\mathrm{V}$ (CB), $V=2.2\,\mathrm{V}$ (SB),  and $V=7.0\,\mathrm{V}$
(conducting regime). The zero-frequency limit gives the Fano factor
$F=S(\omega\!\to\!0)/2e|I|$ which tends to 3 in the SB regime, if we
assume the one lead to be \textit{half-metallic},
$D^{\text{R}}_{\downarrow}/D^{\text{R}}_{\uparrow}\to 0$.
}\label{FIG4}
\end{figure}

Experimentally, the current suppression originating from SB could be
distinguished from other mechanisms by investigating the shot noise of the
system.\cite{Blanter} Figure \ref{FIG4} shows the current noise spectrum for
three different bias voltages representing the CB, SB, and conducting
regimes. The latter is here reached by applying a large negative 
bias, which overcomes the SB by allowing conduction through doubly
occupied states. We here consider the normalized correlation function
$S_{\text{LL}}(\omega)/S_{\text{LL}}(\omega\!\to\!\infty)$ which gives
information about temporal correlations of tunneling events for the left
lead. According to Eq.~(\ref{CT.Somega}), $S_{\text{LL}}(\omega)$ is the
Fourier transform of the current-current correlation function. The constant
part, $S_{\text{LL}}(\omega\to\infty)$,
comes from the autocorrelation contribution,
which is a positive $\delta$-function at vanishing time difference. In the
conducting regime the noise spectrum exhibits a minimum at zero frequency,
since the corresponding contribution to the correlation function is
negative. Whenever the molecule is doubly occupied, further electrons cannot
tunnel in due to the Pauli principle. This is the usual ``antibunching''
effect for fermions. The typical frequency scale apparent in Fig.~\ref{FIG4}
for the conducting case is the inverse of the typical time the molecule
remains doubly occupied, which is of the order of the typical tunneling time
$\tau_0 \equiv (2\pi |t_\alpha|^2
D^{\text{L}}v_{\text{uc}}/\hbar)^{-1}$.\cite{Timm}

In the CB regime, Coulomb repulsion hinders electrons from
entering the molecule and the Fano factor
$F\equiv S(\omega=0)/2e|I|$, which is defined as the ratio of the
zero-frequency noise to the classical Schottky result,\cite{Blanter} is close
to unity. In this case, single electrons tunnel through the system in 
rare, uncorrelated events, and quantum correlations are unimportant.

In contrast, in the SB regime the noise is \emph{enhanced} at
$\omega=0$. This maximum results from a \textit{bunching} of the charge
carriers, which means that several electrons tend to tunnel through the
molecule within a short time interval. On the other hand, the average
waiting time between such events  is comparatively long. Results for the
Fano factor reveal super-Poissonian shot noise, $F>1$, in the SB regime. It
reaches the value $F\approx 3$ for one half-metallic lead,
$D^{\text{R}}_{\downarrow}/ D^{\text{R}}_{\uparrow}\to 0$. This factor can
be understood by taking into consideration that the Fano factor
contains information about the charge of the current-carrying particles and
the quantum correlations between them. In the SB regime, the molecule
is in the
singly occupied state  and has minimal spin for most of the time. Electrons
from the left lead cannot hop onto the molecule until a spin-down electron
is emitted into the right lead. The rate for this process is strongly
suppressed. However, if the spin-down electron does leave the molecule, the
probability for further tunneling processes is high, since electrons of both
spin directions may then tunnel into the molecule. A current is flowing
until a spin-down electron occupies the molecule again, which leads to the
SB state. The Fano factor can be obtained from the following
expression,\cite{Koch2}
\begin{equation}
F = \langle N \rangle \frac{\langle t^2 \rangle - \langle t \rangle^2}
  {\langle t \rangle^2} + \frac{\langle N^2 \rangle - \langle N \rangle^2}
  {\langle N \rangle}. \label{Fanofactor}
\end{equation}
Here $N$ denotes the number of electrons tunneling through the molecule in
one ``bunch'' and $t$ the waiting time between such processes. Since the
molecule is in the same SB state between the events, they are uncorrelated,
giving an exponential distribution of times $t$.
Since the tunneling is dominated
by the three states with $n=0$, $m=-S$ and $n=1$, $m=-S\pm 1/2$, one finds a
probability of $1/2^N$ for $N$ electrons in the bunch. 
This leads to $\langle N\rangle=2$ and $\langle N^2\rangle=6$.
Since $\langle t^2 \rangle = 2\langle t \rangle^2$ for an exponential
distribution, Eq.~(\ref{Fanofactor}) yields $F=3$ for the Fano factor.
This result is of the same fundamental origin as
the super-Poissonian shot noise found for quantum dots
in the SB regime in
Ref.~\onlinecite{Cottet1}, which considers the case $K_2=J=0$ (no local
spin) and nonmagnetic leads. In this case
the SB is induced by application of a magnetic field.\cite{Cottet1}
We note that Eq.~(\ref{Fanofactor}) also gives the observed value $F=1/2$
for the conducting regime.

Bu\l{}ka \textit{et al.}\cite{Bulka} have investigated a similar system, 
consisting of a quantum dot connected to two electrodes,
where the dot and one of the leads are nonmagnetic while the second lead is 
ferromagnetic.
In such an F-N-N junction, a \emph{large} spin can be accumulated
on the dot. This spin accumulation
causes NDC and strongly enhances the current shot noise due to
spin noise activated in the NDC regime.\cite{Bulka}
In contrast, super-Poissonian shot noise in transport through anisotropic
magnetic molecules arises from an interplay of spin-flip processes 
involving a small local spin and spin selection rules.

\begin{figure}[t]
\begin{center}
$\begin{array}{c}
\textbf{(a)} \\
\includegraphics[width=7.5cm,angle=0]{elste5a.eps} \\ 
\textbf{(b)} \\ \\
\hspace{1.0cm} \includegraphics[width=7.0cm,angle=0]{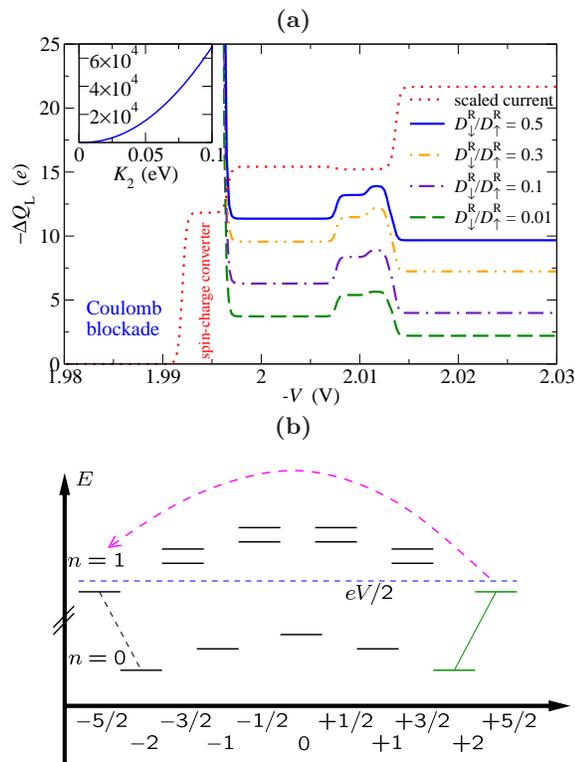} \\
\end{array}$
\caption{(Color online) (a) Excess transmitted 
charge $\Delta Q^{\text{L}}$ as a function of bias voltage in the vicinity of
the first CB step for different polarizations of the right lead:
$D^{\text{R}}_{\downarrow}/D^{\text{R}}_{\uparrow}=0.5,~0.3,~0.1,$ and 0.01.
We assume $J=4\,\mathrm{meV}$, $K_2=1\,\mathrm{meV}$, 
vanishing magnetic field and an initial state with
$n=0$ and $m=2$. The steady-state current for 
$D^{\text{R}}_{\downarrow}/D^{\text{R}}_{\uparrow}=0.1$
is also shown. The inset shows $\Delta Q^{\text{L}}$ 
as a function of the anisotropy constant $K_2$, where 
the bias voltage $V$ corresponds to the arithmetic mean of the first two
excitation energies.
(b) Energy level scheme of the $n=0$ and $n=1$ multiplets.
The molecule is prepared in the initial state with spin $m=S$.}\label{FIG5}
\end{center}
\end{figure}

As noted above, giant spin amplification can occur in transport through
anisotropic magnetic molecules coupled to \textit{nonmagnetic}
leads.\cite{Timm} At bias voltages close to the CB threshold the total spin
transmitted from one lead to the other can become exponentially large at low
temperatures, if the molecule is prepared in a magnetically polarized
initial state at time $t=0$. If one lead is \emph{ferromagnetic}
the steady-state current is highly spin-polarized so that it is necessary
to consider instead the \textit{excess} transmitted spin. Analogously, one
can define the \emph{excess transmitted charge}: The molecule is prepared in
a specific state $|n\rangle$
at time zero and then evolves according to the rate
equations (\ref{rate-equations}). Since it approaches the steady state
exponentially, the excess charge
\begin{equation}
\Delta Q^{\alpha}_n \equiv \int_{0}^{\infty} dt \left[ I^{\alpha}(t)-
  \langle I^{\alpha}\rangle \right]
\end{equation}
is finite. The main observation is that $\Delta Q^{\alpha}_n$ can depend
very strongly on the initial state $|n\rangle$ if the steady state shows SB.
Practically it is much easier to measure the charge accumulation in the
leads instead of the excess spin,\cite{spinaccu} since the accumulated
charge is conserved (except for leakage currents) whereas the spin is not.
Moreover, it should be easier to employ the excess charge for further data
processing.

Results for the excess transmitted charge as a function of bias voltage are
shown in Fig.~\ref{FIG5}(a), where an initial state with $n=0$ and spin
$m=S$ is assumed. The fine structure close to the CB threshold basically
originates from the anisotropy of the local spin. Most striking is the
exponential enhancement of $\Delta Q^{\alpha}_n$ at voltages \textit{above}
the CB threshold, where the steady-state current is already nonzero (but still
small due to SB). When the bias is just large enough to allow the two
transitions from the $n=0$ state with extremal spin to the $n=1$ state with
extremal spin, $m=\pm S\rightarrow \pm (S+1/2)$, the system initially
prepared in a state with \emph{maximal} spin, $m=S+1/2$, for an
exponentially long time only performs transitions between the two extremal
state connected by a solid line in the level scheme in Fig.~\ref{FIG5}(b).
Thus a sizable---and completely spin-polarized---current is flowing until
the molecule overcomes the anisotropy barrier by thermal activation. After
that it rapidly relaxes towards the SB state, which shows a small
steady-state current. On the other hand, if the molecule is prepared in the
state with minimal spin, $m=-S-1/2$, it already starts out in the SB state
and the current is always small. Thus in the first case an
exponentially large excess charge is accumulated in the leads in addition to
the excess spin. In this regime the proposed setup functions as a
\textit{spin-charge converter}, i.e., \emph{spin} information can be read
through a measurement of the excess transmitted \emph{charge},
a much easier task.

On the other hand, if the bias is large enough to allow further transitions
involving other levels, the system rapidly relaxes back towards SB so that
$\Delta Q^{\alpha}_n$ is of order $e$. Note that the excess transmitted
charge increases for increasing anisotropy of the local spin, as can be seen
from the inset in Fig.~\ref{FIG5}(a).

\section{Summary and conclusions}

We have studied transport through a single anisotropic magnetic molecule
coupled to one ferromagnetic and one nonmagnetic lead.
Our main
findings concern the NDC, super-Poissonian noise, and a strong spin dependence
of transmitted charge.

Two types of NDC occur. One appears at \emph{low}
temperatures in an external magnetic field. Here the Coulomb blockade and spin
blockade regimes  are separated by a finite window of bias voltages for which
the current is strongly  enhanced. This is due to the interplay of bias
voltage, magnetic field, and magnetic anisotropy, which allows only transitions
between two specific molecular many-body states and prevents the spin flips
necessary for spin blockade. The other NDC effect appears at \emph{high}
temperatures in the vicinity of the Coulomb blockade threshold. This is a
distinct effect---it can occur even when there is no NDC at low temperatures.
It can be viewed as a manifestation of quantum effects (the Pauli principle and
spin selection rules) at room temperature.

Spin blockade is accompanied by super-Poissonian shot noise, as found earlier
for nonmagnetic quantum dots. Furthermore, the total charge transmitted through
a molecule under a voltage bias depends strongly on its initial spin state in a
certain parameter range. The difference in transmitted charge can diverge
exponentially for low temperatures. This spin-charge conversion presents a
promising method to read out the spin information in molecular-memory
applications.

\acknowledgments

We would like to thank J. Koch, F. von Oppen, C. Romeike, and M. R. Wegewijs
for valuable discussions and the Deutsche Forschungsgemeinschaft 
for financial support through Sonderforschungsbereich 658.


\begin{thebibliography}{99}
\bibitem{rem.superscript}The steady-state current and thus the differential
conductance are identical for the left and right lead due to charge
conservation and we drop the superscript $\alpha$.
\bibitem{rem.maximum}In fact the current maximum does not occur exactly
where $eV/2$ equals the transition energy. Rather, is depends on the
balance between the thermally activated transition rates from $n=0$ to
$1$ and the spin-down transition rates from $n=1$ to $0$ controlled by
$D^{\text{R}}_{\downarrow}/D^{\text{R}}_{\uparrow}$.
\end{thebibliography}
\end{document}